\documentclass{PoS}

\usepackage{inputenc}
\usepackage{graphicx}
\usepackage{epsfig}
\usepackage{color}
\usepackage{longtable}
\usepackage{latexsym}
\usepackage{amsmath}
\usepackage{amsfonts}
\usepackage{amssymb}
\usepackage{dsfont}
\usepackage{verbatim}
\usepackage{bbold}

\usepackage{bm}

\newcommand{\be}{\begin{equation}}
\newcommand{\ee}{\end{equation}}

\title{Baryon resonances coupled to Pion-Nucleon states in lattice QCD}

\ShortTitle{Baryon resonances coupled to Pion-Nucleon states in lattice QCD}

\author{\speaker{Valentina Verduci}\\
        Institute of Physics,  University of Graz, A--8010 Graz, Austria\\
       E-mail: \email{valentina.verduci@uni-graz.at}}

\author{Christian B. Lang\\
        Institute of Physics,  University of Graz, A--8010 Graz, Austria\\
        E-mail: \email{christian.lang@uni-graz.at}}

\abstract{In recent years the study of two particle systems on the lattice has led to excellent results in the meson sector of the QCD spectrum, however baryon resonances mostly remain unexplored. 
We present a study of pion-nucleon systems as decay product of baryon resonances in different channels, with special focus on the nucleon spectrum. We evaluate the correlation functions of single and multi particle interpolators. All the Wick contributions are explicitly computed and the consequences of reduced symmetries in moving frames are taken into account. We discuss the theoretical setup together with results  for $n_f=2$ mass degenerate light quarks.}

\FullConference{The 32nd International Symposium on Lattice Field Theory\\
		 23-28 June, 2014\\
		 Columbia University New York, NY}

\begin{document}

\section{Motivation}
Most of the QCD hadronic spectrum consists of resonances: states with a limited lifetime. They mainly decay under strong interaction processes into many-particle states. Since one of the tasks of lattice QCD is to reproduce the QCD spectrum, the resonant nature of the hadrons has to be taken into account.
In recent years the study of multi-particle systems on the lattice has become possible thanks to new methods for the computation of the correlation matrix, together with improved technical resources and solid analytic approaches for treating the resonances on the lattice. 
While several mesonic channels have been successfully explored, the baryon sector still represents an outstanding challenge. 
In this work we focus on the nucleon sector, which mainly couples to $N\pi$ states, while other channels like $N\pi\pi$ and $N\eta$ are kinematically not accessible for our lattice
setup. In the calculations we explicitly include pion-nucleon interpolators in order to give to the nucleon the possibility of decaying into $N\pi$ states, which seem to couple too weakly with the $qqq$ traditional interpolators.

\section{Method and Setup}
The simulations are performed using configurations with $n_f=2$ mass degenerate dynamical quarks with $m_{\pi}=266$ MeV \cite{Hasenfratz:2008}. The unphysical pion mass leads to an upwards shift in the spectrum by roughly $\sim 130$ MeV.

On our lattice we simulate a pion-nucleon system coupled to the negative (S-wave) and positive (P-wave) parity nucleon sectors, therefore we use both single and multi-particle interpolators, each of them projected onto definite momentum and isospin. In the case of non-rest frame  special attention has to be payed to the discrete symmetry group of the system, which is determined by its total momentum. Each interpolator has to be projected onto the correct irreducible representation. The interpolators used are of the type 
\begin{align}
N^{(i)}(\vec{p}) &= \left\{ \chi_1, \chi_2, \chi_3 \right\} \nonumber \\
\mathcal{O}_{N\pi} (\vec{p}) &= N^{(i)}(\vec {p}) \pi(\vec{0})
\end{align}
where the different $\chi_{(i)}$ differ for their Dirac structure (see  \cite{Lang:2012db}) and couple to different states of the nucleon spectrum. The basis of interpolators is expanded using different distillation smearings, requiring the evaluation of a $9\times 9 $ correlation matrix. 

\begin{figure}[t]\label{}
 \centering
 \includegraphics[scale=0.6]{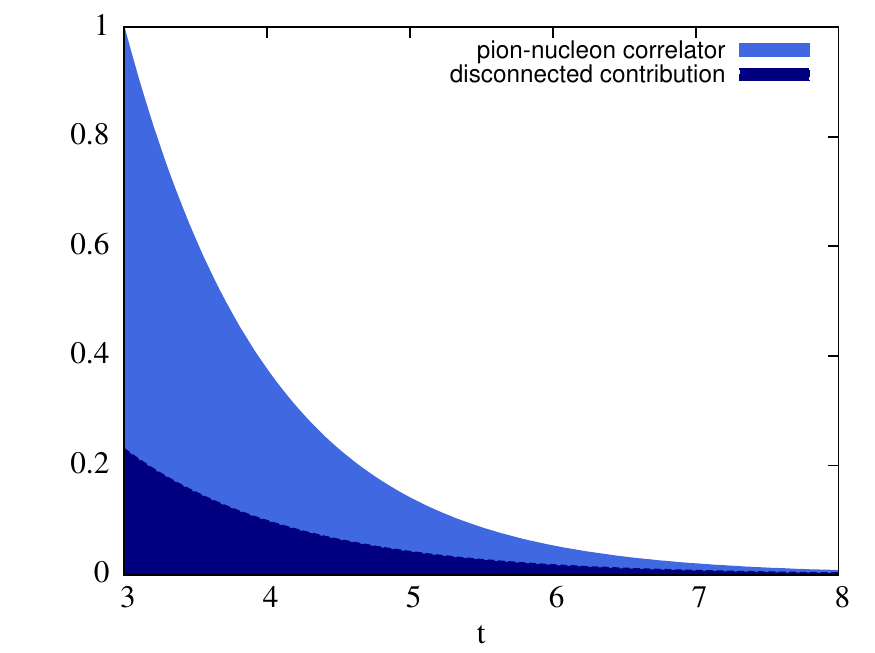} \,\,\,\,\,\,\,\,\,\,\,\,\,\,\,\,\,\,\,\,\,\,\,\,
 \includegraphics[scale=0.35]{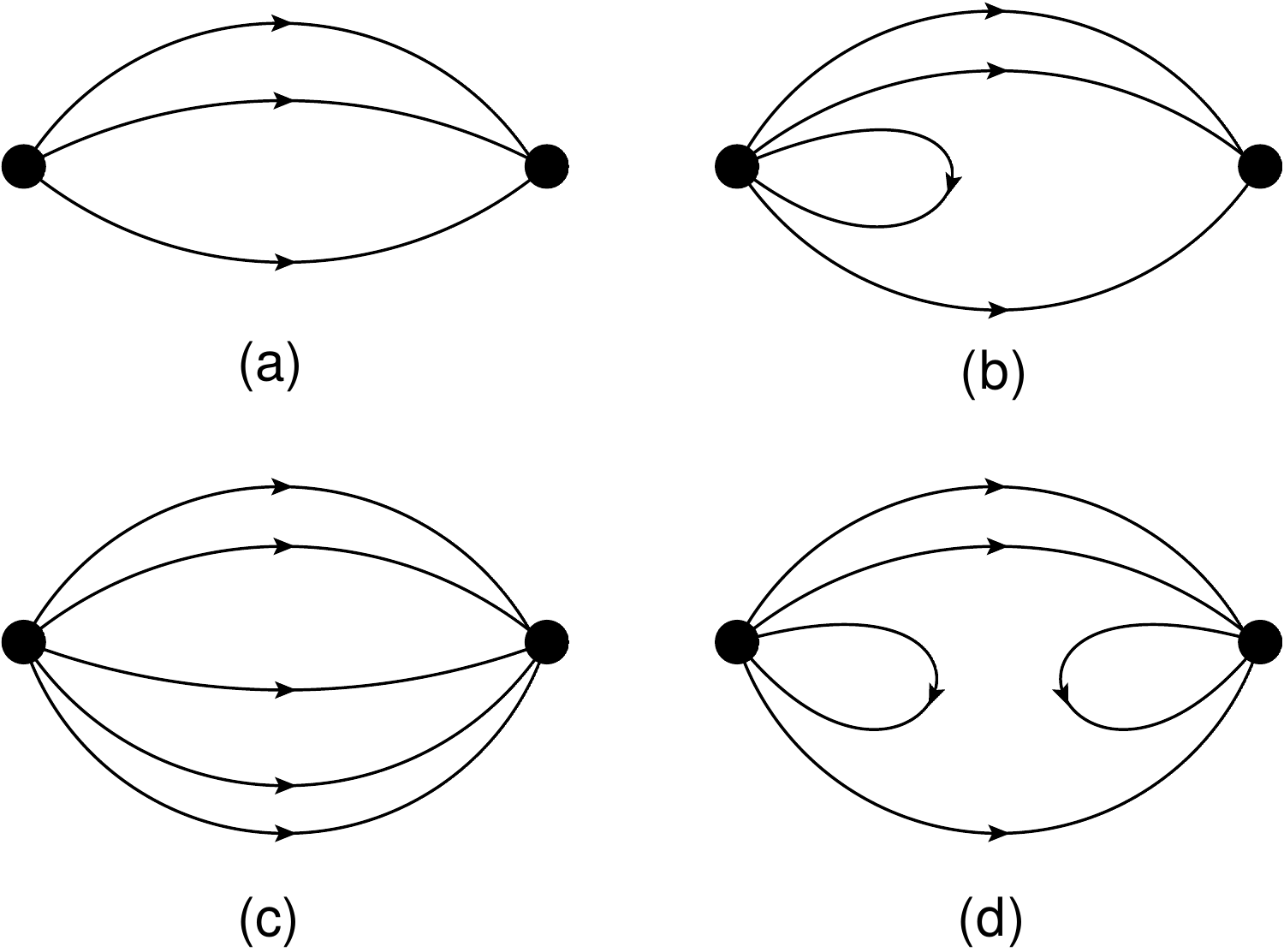}
  \caption{Contribution of the partially disconnected diagrams. Left: the partially disconnected diagrams contribute for $20\%$ of the total diagonal two particle correlator. Right: different types of diagrams included in the calculation.}
\end{figure}

The correlators are computed taking into account the complete set of connected and partially disconnected diagrams which represent all the possible Wick contractions (for details see \cite{Lang:2012db}). The partially connected diagrams in the study of resonances are extremely relevant: in Fig. 1 it is shown that the $20\%$ of the two-particle correlator comes directly from these contributions. Calculations that neglect them are not suited for describing the coupling of a resonance to two particle states since a significant part of the information concerning the interaction and the resonance nature would be lost. The evaluation of the correlation matrix is made possible by the application of the distillation method \cite{Peardon:2009gh}, with perambulators computed in \cite{Lang:2011mn}. The tower of excited states for each channel is obtained using the variational method \cite{Michael:1985ne,Luscher:1990ck}. which also provides the eigenvector composition of each state: its coupling to the different interpolators represents the fingerprint of the energy levels and it will be used for the physical interpretation of the observed spectrum.

\section{S-wave sector}
\begin{figure}[t]\label{Fig:negative}
 \centering
 \includegraphics[width=0.8\textwidth,clip]{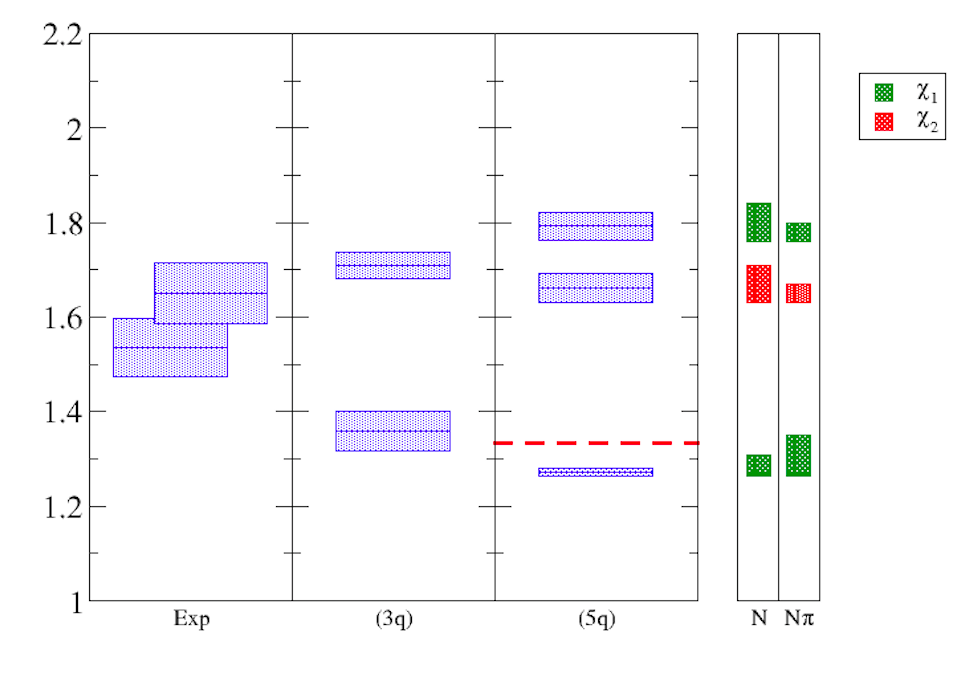}
  \caption[Final results of the pion-nucleon S-wave analysis.]{Results for the S-wave sector of the nucleon spectrum. In the leftmost column we show the expected values (i.e., the physical values of the masses are shifted up of $\Delta E \sim 130$ MeV; the error bars correspond to the experimental width \cite{Beringer:1900zz}). We compare the single particle interpolators analysis (second column from the left) to the new results, where the multi-particle interpolators have been included (third column from the left). The rightmost column represents the eigenvector composition of these last results. }
\end{figure}

The results of the S-wave $N\pi$ system coupled to the negative parity nucleon sector have been published in \cite{Lang:2012db} and are used as benchmark for the physical interpretation of the P-wave sector spectrum.
Note that, differently from the P-wave case (moving frames!), here a definite parity projection can be performed and the negative parity sector of the spectrum can be isolated.
The results are summarized in Fig.\ref{Fig:negative}. When $\mathcal{O}_{N\pi}$ interpolators are included, the effective energy levels show less fluctuations if compared to the pure 3-quark case and a new energy level appears. The eigenvectors composition of the spectrum is well defined when the complete set of interpolators is taken into account (see right column in Fig.\ref{Fig:negative}) and we can conclude that the inclusion of $\mathcal{O}_{N\pi}$ is essential for a reliable picture of the $N_-$ spectrum: a $N\pi$ S-wave state and two single particle states which correspond to $N^*(1535)$ and $N^*(1650)$ are observed.

\section{P-wave sector}
The negative parity sector of the nucleon spectrum couples to a $N\pi$ system in P-wave, therefore non-zero momentum interpolators have to be included in the setup. 
We now present the nucleon spectrum obtained from the study of one-particle and two-particle systems in moving frames, namely $\mathbf{P_{tot}}=(0,0,1)$ and $\mathbf{P_{tot}}=(1,1,1)$ (with a suitable symmetrization in the spatial directions). All the interpolators are projected onto the Irrep $G_1$ of the group ${C_{2v}}$ and $C_{4v}$ (see \cite{Gockeler:2012yj}). 
Other choices are made impossible by the properties of our lattice: in order to hit the energy region which is kinematically interesting for our study, we need the pion to be at rest.
In moving frames a definite parity projection of the interpolators is not possible, therefore we compute the spectrum
using unprojected operators and taking into account that a superposition of the $N_{+}$ 
and $N_-$ spectrum has to be expected when the correlation matrix is diagonalized  \cite{Verduci:2014}. 

\subsection{Single particle spectrum for $\mathbf{P_{tot}}=(0,0,1)$}
\begin{figure}[t]
 \centering
 \includegraphics[width=0.8\textwidth,clip]{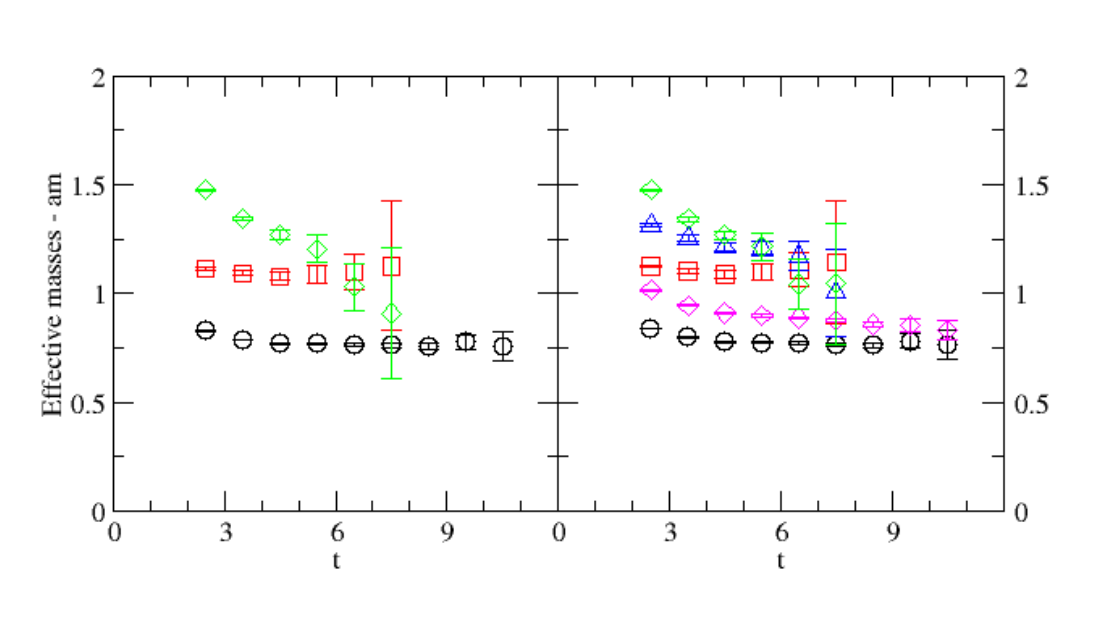}
  \caption[3-quark and 5-quark spectra in moving frame.]{Comparison between the effective masses evaluated in the 3-quark setup and the result of the 5-quark system setup for $\vec{P}_{tot}=(0,0,1)$. All the masses are given in lattice units.
  Note that when the $(4+1)$ quark interpolators are included in the analysis two more states appear: the magenta and the blue state. \label{fig:compare_pwave}}
\end{figure}

Fig. \ref{fig:compare_pwave} (left) shows the nucleon spectrum evaluated when only 3-quark
interpolators are considered. We observe three signals: two of them are stable, while the third signal is quickly decaying and it will not be considered in the further
analysis, but it will serve as reference point for the comparison with the two-particle system. 
The lowest state is without doubt the ground state nucleon. It mainly couples to interpolators of type $N^{(1)}$ and its mass in the 
CMF is $m=1040(6)$ MeV in agreement with the evaluation in $\vec{P}_{tot}=(0,0,0)$  \cite{Lang:2012db}. 
The second state could be either a positive or a negative nucleon excitation, but a comparison with the results of the $\vec{P}_{tot}=\vec{0}$ case allows us to identify it as the fist excited state measured in the negative sector, thanks to its peculiar eigenvector composition.

\subsection{Two-particle spectrum}
When two-particle interpolators are included in the analysis, some unexpected modifications to the spectrum are observed (see Fig. \ref{fig:compare_pwave} right): not only one, but two new levels appear.
The analysis of the eigenvectors composition is crucial for the interpretation of the energy levels (see Fig. 4).

\textbf{1st state}.
The first lower state can be clearly interpreted as the ground state nucleon: it mainly couples to $N^{(1)}$ type interpolators and its mass agrees with the value extracted for $\vec{P}_{tot}=\vec{0}$. 

\begin{figure}[t]
 \centering
 \includegraphics[width=0.82\textwidth,clip]{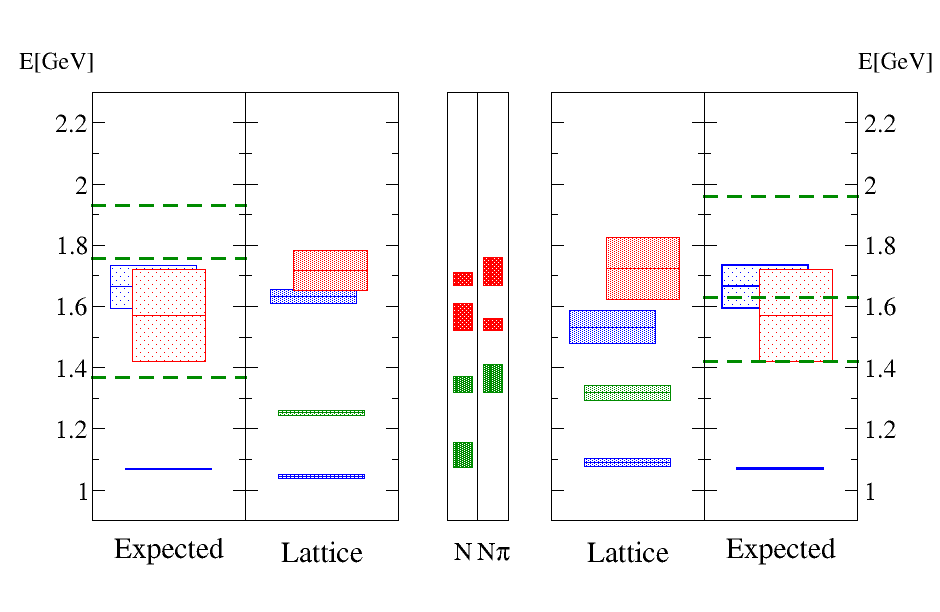}
  \caption[Final results of the pion-nucleon P-wave analysis.]{Results obtained in the CMF on our lattice compared  
  with the expected values (i.e., the physical values of the masses are shifted up of $\Delta E \sim 130$ MeV
  and the error bars correspond to the experimental width \cite{Beringer:1900zz}). The green 
  dashed lines correspond to the expected energy values of the non-interacting $N\pi$ system
  on our lattice: from lower to higher values of the energy we find $\{N(0,0,1)\pi(0)$, \,$N(0)\pi(0,0,1)$, \,$N^*(0,0,1)\pi(0)\}$ for $\vec{P}_{tot}=(0,0,1)$ (left)
  and $\{N(1,1,1)\pi(0)$, \,$N(1,1,0)\pi(0,0,1)$,\, $N^*(1,1,1)\pi(0)\}$ for $\vec{P}_{tot}=(1,1,1)$ (right). Note that in the \textit{expected values} column $N^*(1650)$ has
  not been plotted since our interest is focused in the lower part of the spectrum and from the analysis of the eigenvectors it is clear that none of the energy levels
  plotted in the \textit{Lattice} column can be associated to the resonance.
  The middle column represents the eigenvector composition of each energy level evaluated in our lattice setup (cf. Fig. \ref{Fig:negative}).
 \label{fig:final_plot_pwave}}
\end{figure}

\textbf{2nd state}.
The second state has a very stable signal and its energy can be evaluated with high precision (see Fig. \ref{fig:final_plot_pwave}). Its eigenvector composition leaves no doubts about its nature. This state is clearly a pion-nucleon state: it is dominated by $\pi N^{(1)}$ type interpolators, with an additional small coupling with $N^{(1)}$. This behavior has
already been observed in the S-wave analysis. The state lies below the $N\pi$ theshold, which is not surprising for attractive potential. 

\textbf{3rd state}.
This state corresponds to the first $N^{-}$ state. The main contribution to this level are from $N^{(2)}$ interpolators, with a minor contribution from $\pi N^{(2)}$. The eigenvector composition of this level shows good stability as compared with the 3-quark case. This
is exactly the same signature of the state present in the S-wave study (see Fig. 2, 3rd column, 2nd state). It has to be kept in mind
that, due to the reduced choice in irreducible representations for the interpolators and the absence of a coherent parity projection, a mix of S and P wave in this channel is
expected, therefore both $N^{-}$ and $N^{+}$ states are affected by the coupling with the pion-nucleon system.

\textbf{4th state}.
The most interesting state in our study is the fourth state measured in this $(J,I)$ sector. This state appears when pion-nucleon interpolators are included and it is absent in the
one-particle study of the same problem. From Fig. 4 we see that it mainly couples to $(4+1)$ interpolators of type $\pi N^{(2)}$, plus additional contributions from $N^{(2)}$
interpolators. In order to interpret this state we discuss different hypothesis and
we  make use of the information about the nucleon spectrum which we have collected in our previous study \cite{Lang:2012db}. 

The possibility that this state corresponds to $N^*(1650)$ has to be excluded: in Fig.2 we have seen that $N^*(1650)$ couples to $N^{(1)}$ and mildly to $\pi N^{(1)}$. 
Obviously some fluctuations in the eigenvectors composition could be possible, however a drastic change in the eigenvector signature is not an option, since it represents the fingerprint of the state, as already mentioned.

Another possibility for this state, due to the fact that it couples to two particle interpolators, would be a pion-nucleon state with different momenta than those assigned 
ad hoc in our simulations. In Fig. \ref{fig:final_plot_pwave} we see that the state $N(0)\pi(0,0,1)$ lies in the same energy region  (middle green dashed line in Fig. \ref{fig:final_plot_pwave} leftmost column). This state is a combination of a pion with a ground state
nucleon and such a state cannot couple to $\pi N^{(2)}$ 
interpolators. $N(0)\pi(0,0,1)$ would couple to $\pi N^{(1)}$ interpolators, which is not the case for the state we observe. 

Due to the eigenvector signature of the observed level, it might be argued that this state is the $\pi N^*(1535)$ state, the energy value of which is plotted in Fig. \ref{fig:final_plot_pwave}
(highest green dashed line in the leftmost column). Due to the fact that it is considerably high with respect to our 4th state, we consider it safe to exclude this option.

A possible interpretation for this state is its identification with the Roper resonance. The nature of this state has been discussed in different contexts \cite{Jaffe:2003, Glozman:2004} and
one of the hypothesis suggests a pentaquark nature for this resonance. 
This picture is supported by the fact that the contribution of the $(4+1)$quark interpolators to this state is considerably higher than the one observed in other single particle states and we excluded the possibility for it to be a $N\pi$ state. It has to be noticed  that the  interpolators have no absolute normalization, therefore a statement concerning the dominance of one compared to two particle interpolators cannot be done. Nevertheless a comparison with the other states of the observed spectrum is allowed. This feature would explain the fact that the $N^{+}$ spectrum measured using 3-quark interpolators traditionally does not provide any candidate for this resonance (see, e.g., \cite{Engel:2013}). 

This study has also been performed for a different total momentum: $\vec{P}_{tot}=(1,1,1)$ and the results are plotted in the right hand side of Fig. 4. Due to the instability
of the signal, obtaining a high precision estimation for the different energy levels is beyond the possibility of this study, however it is clear that the emerging picture
is completely compatible with the $\vec{P}_{tot}=(0,0,1)$ study. The states appear with the same ordering and the same eigenvector composition, therefore
all the considerations just made for the other momentum setup hold also in this case. 

The evaluation of the spectrum for two different total momenta is important to exclude inconsistencies in the interpretation of the states. The fact that the states
appear with the same order and the same eigenvector signature supports our conclusions.

\section{Conclusions}
We have computed the pion-nucleon spectrum using nucleon and pion-nucleon interpolators for non zero total momentum. With this choice S-  and P-waves (and therefore positive and negative parity) are mixed and
the interpretation of the evaluated energy levels has to be done with special care. 

We observe, as expected, that including the two-particle interpolators determines the presence of a $\pi N$ state with a clear and stable signal. However also the rest of the spectrum is subject 
to a major modification: a new state appears. This state strongly couples to the (4+1) quark interpolators, but it seems not to be compatible with any 2-particle state. 
A possible interpretation is identifying this state with the Roper resonance. The energy level, however, still appears above
 the state identified as the $N^*(1535)$. It is possible that an extrapolation to physical masses leads to the inversion of the mass ordering, however this cannot
be done in the context of the current study. \vspace{12pt}

\textit{Acknowledgements:}
Support from the Austrian Science Fund (FWF) through the grant DK W1203-N16 
is gratefully acknowledged.

\end{document}